\documentclass[twocolumn,floatfix,pre,showpacs]{revtex4-1}
\usepackage{graphicx}
\usepackage{amssymb}
\usepackage{amsmath}
\usepackage{color}
\usepackage{psfrag}
\usepackage{epsfig}
\usepackage{bbm}
\usepackage{bm}
\usepackage{dsfont}
\usepackage[capitalize]{cleveref}
\usepackage{csquotes}

\newcommand{\beq}{\begin{eqnarray}}
\newcommand{\eeq}{\end{eqnarray}}

\newcommand{\eq}[1]{Eq.~(\ref{#1})}
\newcommand{\seq}[1]{$\left[ \text{see~Eq.~(\ref{#1})} \right]$}

\newcommand{\fig}[1]{Fig.~\ref{#1}}

\begin{document}
\title{Non-equilibrium dynamics of a confined colloidal bilayer in planar shear flow.}
\author{Tarlan A. Vezirov and Sabine H.~L.~Klapp}
\affiliation{
  Institut f\"ur Theoretische Physik,
  Hardenbergstr. 36,
  Technische Universit\"at Berlin,
  D-10623 Berlin,
  Germany
}

\date{\today}
\begin{abstract}
Using Brownian dynamics (BD) simulations we investigate the impact of shear flow on structural and dynamical properties of a system of charged colloids confined to a narrow slit pore. Our model consists of spherical microions interacting through a Derjaguin-Landau-Verwey-Overbeek (DLVO) and a soft-sphere potential. The DLVO parameters were chosen according to a system of moderately charged silica particles (with valence $Z\sim$35) in a solvent of low ionic strength. At the confinement conditions considered, the colloids form two well-pronounced layers. In the present study we investigate shear-induced transitions of the translational order and dynamics in the layers, including a discussion of the translational diffusion. In particular, we show that diffusion in the shear-melted state can be described by an analytical model involving a single shear-driven particle in a harmonic trap. We also explore the emergence of zig-zag motion characterized by spatio-temporal oscillations of the particles in the layers in 
the vorticity direction. Similar behavior has been recently observed in experiments of much thicker colloidal films. 
\end{abstract}
\pacs{{\tt 82.70.Dd,83.50.Ax,05.70.Ln,82.70.Kj}}
\maketitle
%%%%%%%%%%%%%%%%%%%%%%%%%%%%%%%%%%%%%%%%%%%%%%%%%%%%%%%%%%%%%%%%%%%%%%%%%

%{\tt TO DO}

%{\tt RG please}
%\begin{itemize}
%  \item please plot fig 2 consistent with the rest.
%\end{itemize}

%%%%%%%%%%%%%%%%%%%%%%%%%%%%%%%%%%%%%%%%%%%%%%%%%%%%%%%%%%%%%%%%%
%%%%%%%%%%%%%%%%%%%%%%%        INTRO        %%%%%%%%%%%%%%%%%%%%%
%%%%%%%%%%%%%%%%%%%%%%%%%%%%%%%%%%%%%%%%%%%%%%%%%%%%%%%%%%%%%%%%%
\section{Introduction \label{SEC:INTRO}}

Colloidal systems subject to one or several confining surfaces have been a focus of intense theoretical and experimental research for decades \cite{Schoen07}. Recently, interest in this area, which has long been devoted to the structural and phase behavior of confined colloids, has shifted towards non-equilibrium phenomena arising in presence of a driving mechanism, examples being shear flow, magnetic field gradients or electric fields \cite{Loewen10, Glanz12}. \par
The reasons for the strong interest in driven, confined colloidal systems are manifold: 
First, from the experimental side, confined colloidal suspensions can serve as well controllable and (compared to atomic systems) accessible model systems to study the behavior of condensed matter in restricted geometries. An important aspect, especially for the investigation of the dynamics, is that colloidal particles are typically large (and thus slow) enough to be followed by real-space methods such as video microscopy. Two recent examples in the context of transport phenomena are the observation of excitons (kinks and antikinks) in driven colloidal monolayers \cite{Bechinger12}, and the crossover from single-file to conventional Fick diffusion in colloidal nano channels \cite{Siems12}. Second, from the theoretical and simulation side, many properties of colloids in and out of equilibrium can be described by established {\it effective} interaction models where the solvent enters only implicitly. This enables one to directly employ a broad variety of semi-analytical methods such as density functional 
theory \cite{Evans79} and its dynamical extensions \cite{Tarazona00}, mode-coupling theory (MCT) \cite{Brader07, Brader10}, as well as particle-resolved computer simulations
such as Brownian Dynamics (BD).  For example, strong interest has been recently devoted to the understanding of the nonlinear response of single colloids driven through a highly correlated colloidal environment ("active microrheology"), a topic which has been studied by simulations, MCT, and by experiment \cite{Winter12, Gazuz09}. Finally, driven colloidal systems have a variety of applications such as in lubrication science \cite{Ning12, Wongy10}, in some medical contexts (i.e., synovial fluid lubrication, tissue shearing) \cite{Jin11, Tan13, Nicollea12, Mitchell13, Leslie13} but also in macroscopic contexts such as traffic jam \cite{Helbing01, Champagne10}.  \par
Among the broad variety of driving mechanisms, the present paper is devoted to the impact of shear. In particular, we present a detailed BD simulation study
of a system of charged colloidal spheres under the combined influence of strong spatial confinement, as realized by two plane-parallel, smooth walls, and planar, steady shear flow. We focus on dense systems composed of only two layers of particles between the confining plates, i.e., so-called "bilayers". Our main goal is to explore the dynamical features of the particles accompanying the structural changes induced by shear. Indeed, previous simulation studies have already shown that application of shear on dense colloidal bilayers can induce complex changes of the in-plane positional structure \cite{Messina06}. One key aspect of the present study concerns the associated changes of the translational mobility as measured by the mean-squared displacement (MSD). In particular, we show that the combined impact of confining geometry and strong correlations has a profound impact on the in-plane MSD, which strongly differs from that observed for a free particle in shear flow. We also suggest a corresponding 
analytical model. Another key question of our study is whether bilayer systems display shear-induced collective, oscillatory modes. Indeed, such dynamical modes (a "zig-zag"-like motion in vorticity direction) were recently observed in an experimental study of three-dimensional (3D) colloidal crystals under shear \cite{Derks09}. Here we demonstrate that such oscillatory motion (in steady shear) also occurs in strongly confined systems, with characteristics very similar to those observed in the experiment. \par
In our computational study we concentrate on bilayer systems whose zero-shear structure is characterized by crystalline in-plane order with square-like symmetry. Indeed, it is well known that already the equilibrium  behavior of strongly confined colloidal systems is quite complex as they can form a variety of crystalline states depending on the details of the wall separation, density, and interaction range \cite{Oguz12}. In the present case, the interactions are described on an effective (i.e., solvent-free) level via Derjaguin-Landau-Verwey-Overbeek (DLVO) theory, with parameters suitable for moderately charged silica nanoparticles. The resulting model has already been successfully applied in previous studies \cite{Grandner09}. In particular, we have shown that the charged-colloid bilayer forms square, hexagonal and buckled phases, in qualitative agreement with the well-known behavior of hard spheres \cite{Grandner08}. Thus, the present model seems to be an ideal candidate to investigate the dynamical behavior of 
sheared bilayers. Moreover, by working with an already tested parameter set \cite{Grandner09,Klapp08} we can ensure that our predictions can, in principle, be tested in an experiment.\par
In the subsequent section~\ref{SEC:MODS}, we describe our model system and the details of our computer simulations. In the present work we follow earlier studies of shear-driven suspensions \cite{Besseling12, Wilms12} in that we neglect hydrodynamic interactions, an approximation which strongly facilitates
the computations in terms of efficiency. Numerical results are presented in Sec.~\ref{SEC:SIMD}. We start in Sec.~\ref{SEC:SIPT} by a discussion of shear-induced structural changes, followed by a numerical analysis of the translational single-particle dynamics in Sec.~\ref{SEC:TM}. In Sec.~\ref{SEC:DT} we suggest an analytical model which successfully describes the simulation results for the MSD at intermediate shear rates. Finally, in Sec.~\ref{SEC:ZM} we report our observations on zig-zag motion at high shear rates. We close with a summary and conclusion in Sec.~\ref{SEC:CONC}.

%%%%%%%%%%%%%%%%%%%%%%%%%%%%%%%%%%%%%%%%%%%%%%%%%%%%%%%%%%%%%%%%%
%%%%%%%%%%%%%%%%%%%        Model System     %%%%%%%%%%%%%%%%%%%%
%%%%%%%%%%%%%%%%%%%%%%%%%%%%%%%%%%%%%%%%%%%%%%%%%%%%%%%%%%%%%%%%%
\section{Model system \label{SEC:MODS}}
We consider a colloidal suspension composed of (silica) macroions, salt ions, counterions, and solvent molecules. In order to make this complex, multicomponent, system accessible for a theoretical study we employ the DLVO approximation. In this framework, the focus is on the macroions, whereas the influence of the individual counter- and salt ions is described in a mean-field manner \cite{Hansen}. The result of such an approximation is a screened Coulomb interaction between the macroions, i.e., a Yukawa-like potential
\beq
  U_{DLVO}(r) = W\frac{\exp\left(-\kappa r\right)}{r},
 \label{UDLVO}
\eeq
with a prefactor
\beq
  W = \frac{\left({\tilde Z} e_{0}\right)^{2}}{4 \pi \epsilon_{0} \epsilon}  \exp\left(\kappa\sigma\right).
\label{W}
\eeq
As seen from Eq.~\eqref{W}, the prefactor $W$ is a function of the elementary charge $e_{0}$, the permittivity of the vacuum $\epsilon_{0}$, the permittivity of the solvent $\epsilon$, the effective valency ${\tilde Z}=Z/(1+\kappa \sigma /2)$ and the inverse Debye screening length 
\beq
  \kappa = \sqrt{\frac{e_{0}^{2}}{\epsilon_{0}\epsilon k_{B} T}\left(Z\rho+2IN_{A}\right)}.
\label{kappa}
\eeq
The inverse Debye length determines the range of repulsion between the macroions. In Eq.~\eqref{kappa}, $\rho$ is the density of macroions, $T$ is the temperature of the system and $I$ is the ionic strength which measures the salt concentration in the solution. Finally, $N_{A}$ and $k_{B}$ are the Avogadro constant and the Boltzmann constant, respectively. \par
In addition to the charge-induced interaction we also take into account the steric repulsion between macroions. To this end we supplement the DLVO potential by the repulsive part of the Lennard-Jones potential, that is, the soft sphere (SS) interaction
\beq
 U_{SS}(r) = 4\epsilon_{SS}\left(\sigma/r\right)^{12},
\label{USS}
\eeq
with $\epsilon_{SS}/k_{B}T=$1.
So the total fluid-fluid interaction reads
\beq
U_{FF}(r)=U_{DLVO}(r)+U_{SS}(r).
\label{UFFTILDE}
\eeq
Following earlier work \cite{Smit01} we have truncated and shifted the potential at the cutoff radius $r_{c}$=2.97$\sigma$, such that $U_{FF}(r_{c})$=0 and $\left.\mathbf{F}_{FF}(\mathbf{r})\right\vert_{|\mathbf{r}|=r_{c}}$=0 with $\mathbf{F}_{FF}(\mathbf{r})=-\mathbf{\nabla}U_{FF}(r)$.
The parameters for the DLVO interaction were set in accordance to the experimental setups in \cite{Grandner09, Klapp08} to $Z$=35, $\sigma$=26 nm, $T$=298 K, $\epsilon$=78.5 and $I$=10$^{-5}$ mol/l. These parameters determine the dimensionless prefactor of the Yukawa potential, $W^{*}=W/(k_{B}T\sigma)$, and the inverse Debye screening length $\kappa^{*}=\kappa\sigma$ via Eqs.~\eqref{W} and \eqref{kappa}. In the present study we have chosen the reduced density $\rho^{*}=\rho\sigma^{3}$=0.85, yielding $W^{*}$=123.366 and $\kappa^{*}$=3.22.\par
In using DLVO theory we have to keep in mind that the standard DLVO potential introduced in Eqs.~\eqref{UDLVO}-\eqref{kappa} pertains to a bulk system, where each macroion is surrounded by a spherical cloud of counterions. Here, we are considering strongly confined systems where the counterion clouds are distorted. Nevertheless, as we have already seen in previous investigations \cite{Grandner09, Klapp08}, the bulk DLVO theory yields a good description at least of structural phenomena such as layering. This can be explained by the fact that, at the conditions considered ($\kappa^{*}=3.22$), the thickness of the screening layer $\kappa^{-1}$ is small compared to the colloidal diameter. As a consequence the deformation of the counterion clouds is only a minor effect. Here we use similar conditions as in \cite{Grandner09, Klapp08} and thus employ bulk DLVO theory.\par
Finally, to mimic a slit-pore geometry, the colloidal particles are confined by two soft plane-parallel walls of infinite extent in the $x-y$ plane and located at $z=\pm L_{z}$/2. The interaction between the colloids and the walls is described by 
\begin{align}
U_{FS}(z) &=\frac{4 \pi \epsilon_{W}}{5}\left(\left(\frac{\sigma}{L_{z}/2-z}\right)^9+\left(\frac{\sigma}{L_{z}/2+z}\right)^9 \right),
\label{UFS}
\end{align}
with $\epsilon_{W}/k_{B}T=$1.
%%%%%%%%%%%%%%%%%%%%%%%%%%%%%%%%%%%%%%%%%%%%%%%%%%%%%%%%%%%%%%%%%
%%%%%%%%%%%%%%%%%%%%%%      Simulation details      %%%%%%%%%%%%%%%%%%%%%
%%%%%%%%%%%%%%%%%%%%%%%%%%%%%%%%%%%%%%%%%%%%%%%%%%%%%%%%%%%%%%%%%
\section{Simulation details \label{SEC:SIMD}}
\begin{figure}
\includegraphics[width=\linewidth]{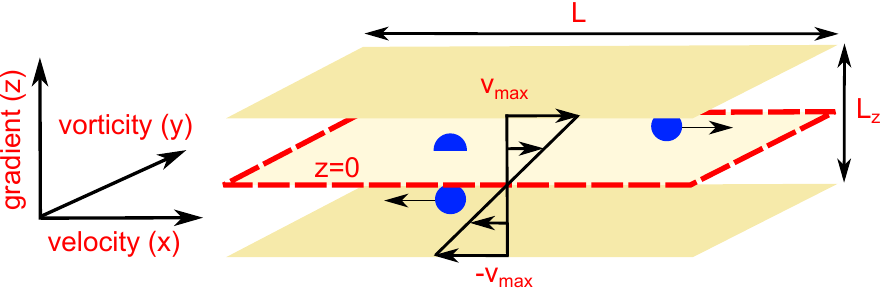}
\caption{(Color online) Definition of the coordinate system and schematic illustration of the shear cell. The plane with red dashed edges indicates the zero-velocity plane, which is positioned at $z$=0.}
\label{FIG:Sketch}
\end{figure}
To investigate the non-equilibrium properties of our model system of confined colloidal particles we employ overdamped Brownian dynamics (BD) simulations, using an algorithm suggested by Ermak \cite{Ermak75}. In this framework, the equation of motion reads
\begin{align}
\mathbf{r}_{i}(t+\delta t) = \mathbf{r}_{i}(t)+\frac{D_{0}}{k_{B} T}\mathbf{F}_{i}(t)\delta t+\delta \mathbf{W}_{i}+\dot{\gamma}z_{i}(t)\delta t \mathbf{e}_{x},
\label{EQ:Eqmot}
\end{align}
where $\mathbf{r}_{i}(t)=\left(x_{i}(t),y_{i}(t),z_{i}(t)\right)$ describes the position of the $i$-th particle at time t, and $ \mathbf{F}_{i}(t)$ is the sum of conservative forces resulting from the pair interaction between particles \seq{UFFTILDE} and the repulsive interaction between the particles and the wall \seq{UFS}. We apply a plane Couette shear flow in $x$-direction with gradient in $z$-direction and vorticity in $y$-direction. As reflected by the last term on the right side of \eq{EQ:Eqmot}, we {\it impose} a linear velocity profile characterized by the shear rate $\dot{\gamma}$. This is a common approximation also employed in other recent simulation studies of sheared colloidal systems (see e.g. \cite{Besseling12, Messina06}). As we will later see, however, the effective shear rate can indeed differ from $\dot{\gamma}$. The shear geometry with the zero-velocity plane at $z$=0 is illustrated in \fig{FIG:Sketch}.\par
Within the framework of BD, the influence of the solvent on each colloidal particle is mimicked by a friction constant and a random Gaussian displacement $\delta \mathbf{W}_{i}$. The friction constant is set to $(D_{0}/k_{B} T)^{-1}$, where $D_{0}$ is the short-time diffusion coefficient. The value $\delta \mathbf{W}_{i}$ has zero mean and variance $2D_{0}\delta t$ for each Cartesian component. The timescale of the system was set to $\tau=\sigma^{2}/D_{0}$, which defines the so-called Brownian time. The timestep was set to $\delta t=10^{-5}\tau$. We consider a system of 1058 particles which are confined in a rectangular $L\times L\times L_{z}$ box. Periodic boundary conditions in flow ($x$) and vorticity ($y$) direction were applied. The side lengths of the shear cell were set to $L=23.786\sigma$ and $L_{z}=2.2\sigma$, respectively, resulting in a reduced density of $\rho^{*}$=0.85. In previous Monte Carlo simulations \cite{Grandner08} it was already shown that under these confinement conditions the 
equilibrium structure of the system corresponds to two layers of particles with square-like positional order. In the present work we begin our simulations from this initial configuration, which is illustrated in \fig{FIG:Snapshot}a).
%%%%%%%%%%%%%%%%%%%%%%%%%%%%%%%%%%%%%%%%%%%%%%%%%%%%%%%%%%%%%%%%%
%%%%%%%%%%%%%%%%%%%%%%%      Results    %%%%%%%%%%%%%%%%%%%%%
%%%%%%%%%%%%%%%%%%%%%%%%%%%%%%%%%%%%%%%%%%%%%%%%%%%%%%%%%%%%%%%%%
\section{Results \label{SEC:RES}}
\subsection{Shear-induced structural changes \label{SEC:SIPT}}
As a starting point we investigate the shear-induced changes of the structure of the confined colloidal bilayer.
\begin{figure}
\includegraphics[width=0.9\linewidth]{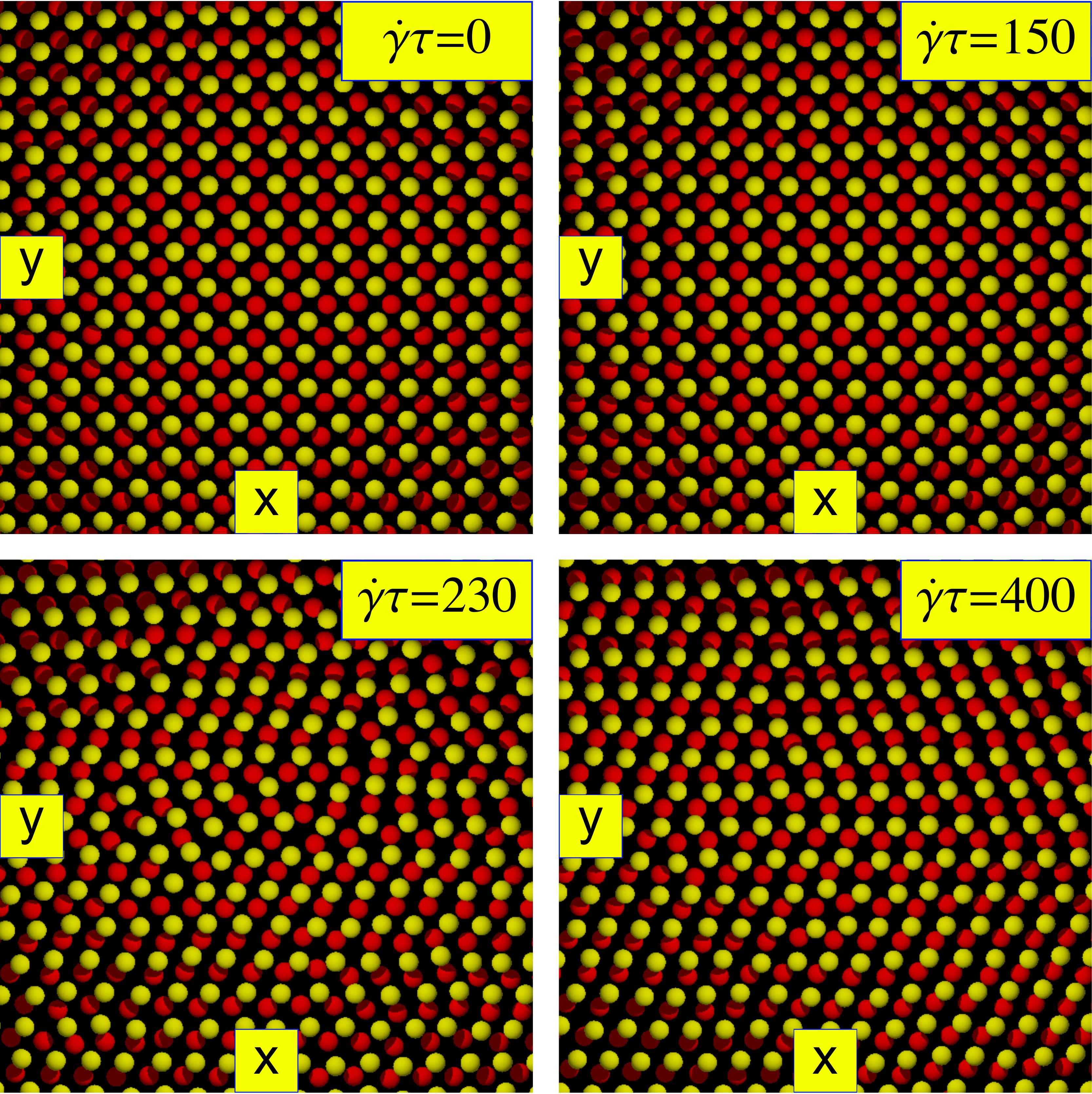}
\caption{(Color online) Simulation snapshots at $\rho^{*}$=0.85 and $L_{z}$=2.2$\sigma$ for different shear rates. The yellow (red) circles represent particles of the upper (lower) layer.}
\label{FIG:Snapshot}
\end{figure}
\begin{figure}
\includegraphics[width=1\linewidth]{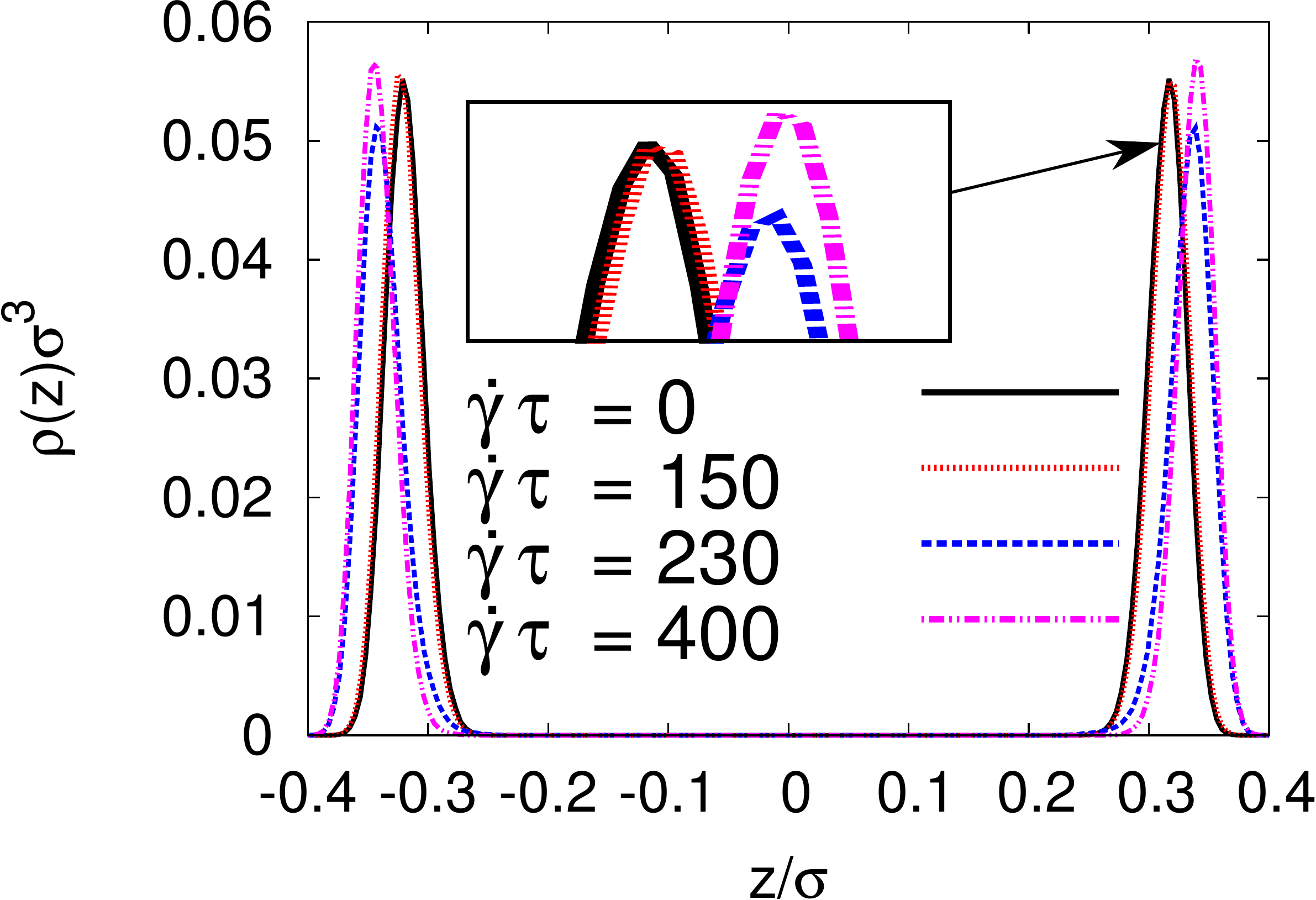}
\caption{(Color online) Density profiles in the shear gradient (and confinement) direction for different shear rates.}
\label{FIG:Density}
\end{figure}
\begin{figure}
\includegraphics[width=\linewidth]{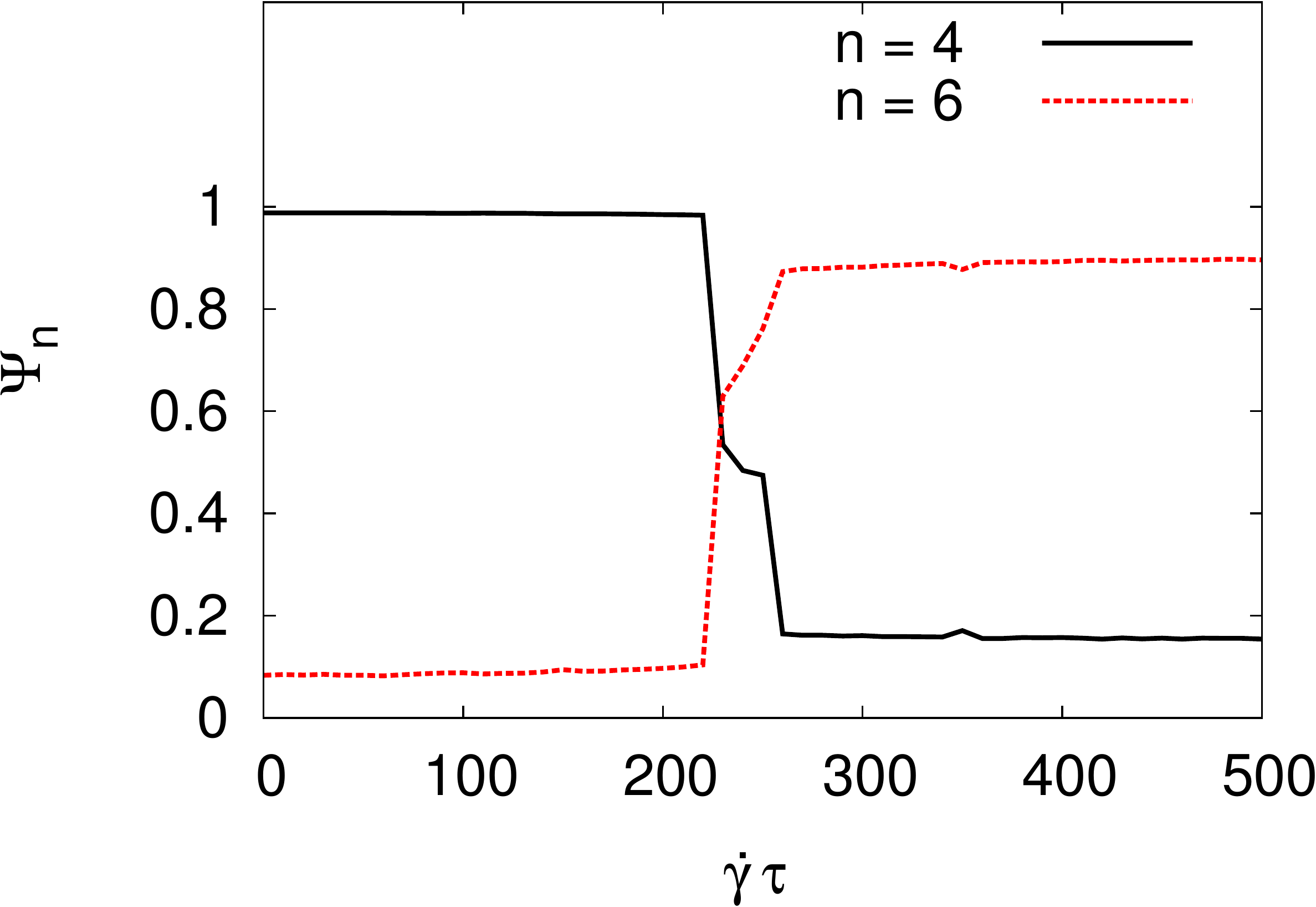}
\caption{(Color online) In-plane order parameter for square ($\Psi_{4}$) and hexagonal ($\Psi_{6}$) symmetry as a function of the shear rate $\dot{\gamma}$.}
\label{FIG:Order}
\end{figure}
The system was equilibrated for $5\times10^6$ steps (i.e. 50$\tau$), after which the shear force was switched on. The simulations were then carried on for an additional period of 50$\tau$, during which a non-equilibrium steady state was reached. Only then we started to analyze the system properties.\par
As a first step, we calculated the density profile in z-direction, $\rho(z)=\langle N(z)/N\rangle$ with $N(z)$ being the instantaneous number of particles at a given vertical distance $z$. Second, we have obtained the translational in-plane order parameters $\Psi_{n}$ defined by 
\begin{align}
\Psi_{n}=\left\langle \frac{1}{N_{l}}\sum_{i=1}^{N_{l}} \frac{1}{N_{i}^{b}} \bigg| \sum_{j=1}^{N_{i}^{b}}\exp(in\theta_{j}) \bigg| \right\rangle.
\label{EQ:Order}
\end{align}
In \eq{EQ:Order}, $N_{l}$ is the instantaneous number of particles in the layer considered [the position of this layer can be extracted from $\rho(z)$], and $N_{i}^{b}$ is the number of neighbors of particle $i$ in the layer. The number is calculated from the intralayer pair correlation function, $g_{intra}(R)$, where $R=\sqrt{x_{ij}^{2}+y_{ij}^{2}}$ is in the in-plane separation of two particles. The intralayer correlation function is defined as
\begin{align}
g_{intra}(R)=\frac{L^{2}}{N_{l}^{2}A} \left\langle \sum_{i=1}^{N_{l}}\sum_{j\neq i}^{N_{l}} \delta(\mathbf{R}-\mathbf{R}_{ij}) \right\rangle,
\label{EQ:Gintra}
\end{align}
where $A$ is the corresponding annulus. The number $N_{i}^{b}$ then follows as the instantaneous number of neighbors of particle $i$ within a Radius $R_{ij}^{c}$, corresponding to the first minimum in $g_{intra}(R)$. 
In the following we focus on the order parameters $\Psi_{4}$ and $\Psi_{6}$ measuring how close the system is to a perfect square ($\Psi_{4}$=1, $\Psi_{6}$=0) or hexagonal order ($\Psi_{4}$=0, $\Psi_{6}$=1), respectively.\par
\begin{figure}
\includegraphics[width=\linewidth]{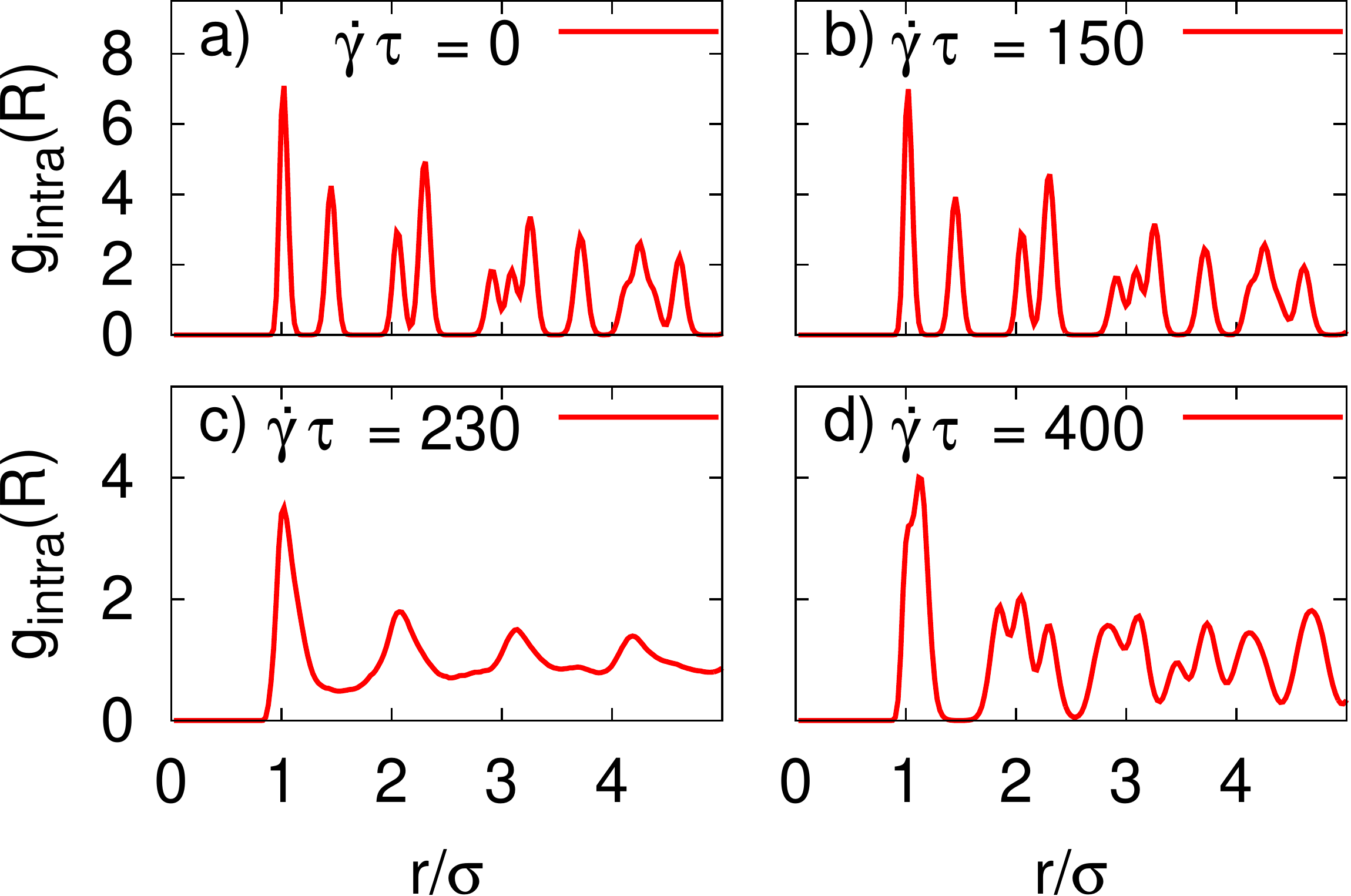}
\caption{(Color online) Intralayer distribution function $g_{intra}(R)$ at different shear rates.}
\label{FIG:Intra}
\end{figure}

Numerical results for $\rho(z)$, $\Psi_{4/6}$ and $g_{intra}(R)$ are plotted in Figs.~\ref{FIG:Density},~\ref{FIG:Order} and~\ref{FIG:Intra}, respectively. For their interpretation, it is instructive to inspect additionally the simulation \enquote{snapshots} at different shear rates shown in Figs.~\ref{FIG:Snapshot}a)-d). From these snapshots and the plots of $\Psi_{n}$ we find that there are three different structural regimes. At small shear rates ($\dot{\gamma}\tau\!\!\leq\!\!220$) the system retains in-plane crystalline order with square symmetry. Indeed, as seen from  Fig.~\ref{FIG:Order}, the order parameter $\Psi_{4}$ is close to 1 throughout this range of shear rates. In the subsequent, relatively small regime ($230\!\leq\!\dot{\gamma}\tau\!\leq\!250$) the intralayer structure becomes disordered [see~Fig.~\ref{FIG:Snapshot}c)], and the system can be considered as shear-melted. Finally, at ($\dot{\gamma}\tau\!\!\geq\!\!260$) the system displays a reentrant crystallization into a state with hexagonal 
symmetry. This is characterized by an increase of the hexagonal order parameter $\Psi_{6}$ above the threshold value 0.7 [see~Fig.~\ref{FIG:Order}]. \par
The shear-induced structural changes are also reflected by the density profile [see~Fig.~\ref{FIG:Density}]. At small shear rates, at which the system persists in the square-crystalline state the layers remain to have the same distance with respect to each other (as inferred from the location of the density peaks) as at $\dot{\gamma}\tau\!=\!0$. But entering the shear-melted state, the distance between two layers increases. This can be explained by the fact that the shear force now yields a net motion of the layers (see~Sec.~\ref{SEC:TM}), while at the same time, particles can now escape their lattice positions. This induces an effective repulsion between the layers, which the system compensates by accepting a smaller distance between the layers and the confining walls. Finally, we briefly inspect the correlation function $g_{intra}(R)$ plotted in Fig.~\ref{FIG:Intra}. It is seen that this function displays profound changes when going from one shear regime to the next. Specially, in the range $\dot{\gamma}\
tau\!\!\leq\!\!220$, $g_{intra}(R)$ has a well defined form corresponding to the 4-fold symmetry [see~Fig.~\ref{FIG:Intra}a)-b)]. Entering the disordered state, the intralayer distribution function displays a liquid-like structure reflecting a loss of long-range positional correlations. At even larger shear rates $g_{intra}(R)$ becomes again crystal-like, this time reflecting a 6-fold symmetry.\par
Similar results were observed in an earlier simulation study of a colloidal bilayer under shear \cite{Messina06}. However, the calculations in \cite{Messina06} were performed at a significantly lower reduced density. This may explain why the first peak of $g_{intra}(R)$ in the hexagonal state is a single peak in \cite{Messina06}, whereas it is splitted in our study [see~Fig.~\ref{FIG:Intra}d)]. We interpret this splitting as an indication that the shear slightly distorts the hexagonal symmetry in very dense layers. To this end we view the hexagonal lattice as a collection of strings along the flow (i.e., $x$-) direction. Due to the shear and the presence of an adjacent layer [see~Fig.~\ref{FIG:Snapshot}d)], the distance between nearest neighbors in different strings along the flow (i.e., $x$-) direction will be slightly larger than that of nearest neighbors in the same string. It seems plausible that this effect will be partially pronounced at high densities such as the one considered in our study ($\rho^{*}
=085$). 
\subsection{Translational dynamics \label{SEC:TM}}
\begin{figure}
\includegraphics[width=\linewidth]{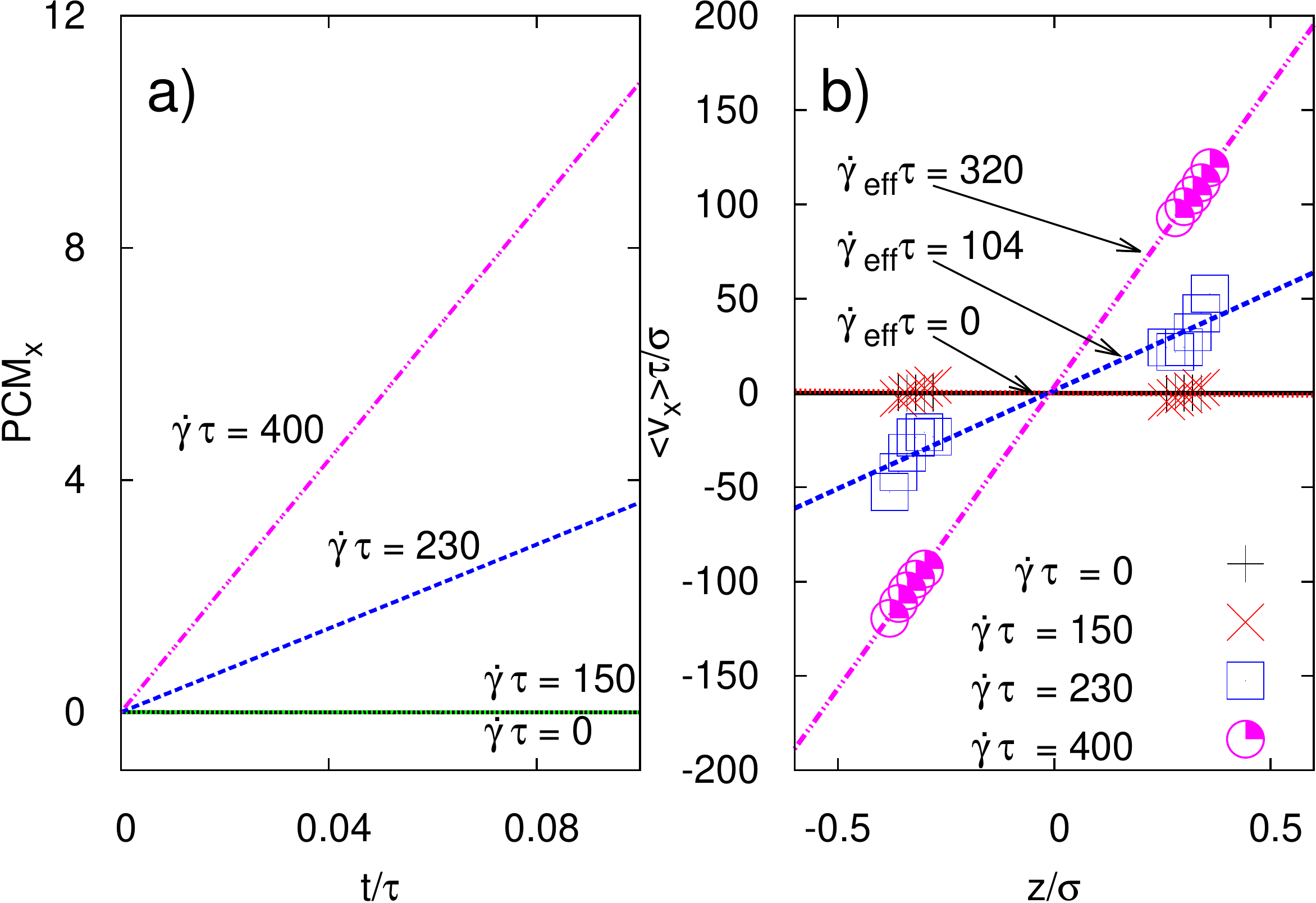}
\caption{(Color online) The position of the center of mass (PCM) in flow direction (a) and the velocity profile of the system (b) at different shear rates.}
\label{FIG:Velocity}
\end{figure}

We now turn to the discussion of the dynamical properties of our bilayer system. We start by investigating the motion of one of the layers in the shear flow. To this end we calculate the position of the layer's center of mass,
\begin{align}
\Delta\mathbf{r}_{cm}(t)=\left\langle\frac{1}{N_{l}}\sum_{i=1}^{N_{l}}\left(\mathbf{r}_{i}(t+t')-\mathbf{r}_{i}(t')\right)\right\rangle,
\label{EQ:PCM}
\end{align}
with $t'$ referring to the starting point of the calculation.
In Figure~\ref{FIG:Velocity}a) we plot results for the function $\Delta\mathbf{r}_{cm}(t)$ in flow direction. It is seen that the square-ordered structure appearing at small shear rates is characterized by zero net motion of the layers. This changes only when $\dot{\gamma}$ is increased towards values pertaining to the shear-melted state. Finally, after the reentrance of the crystalline state (with hexagonal order) we observe in Fig.~\ref{FIG:Velocity}a) even faster net motion of the layers. These trends are also reflected by the velocity profile defined as 
\begin{align}
v_{x}(z)=\frac{1}{N(z)}\sum_{i=1}^{N(z)}\frac{x_{i}(t+\Delta t)-x_{i}(t)}{\Delta t},
\label{EQ:Velprof}
\end{align}
with $N(z)$ corresponding to the number of particles at the considered $z$-position.
Results for $v_{x}(z)$ are plotted in Fig.~\ref{FIG:Velocity}b). The data are strongly accumulated at $z\approx \pm 0.4\sigma$, corresponding to the peaks in the density profile [see~Fig.~\ref{FIG:Density}]. Nevertheless, as seen in Fig.~\ref{FIG:Velocity}b), we can fit the data, to a reasonable degree, by linear functions, consistent with our ansatz for the imposed shear flow in \eq{EQ:Eqmot}.
\begin{figure}
\includegraphics[width=\linewidth]{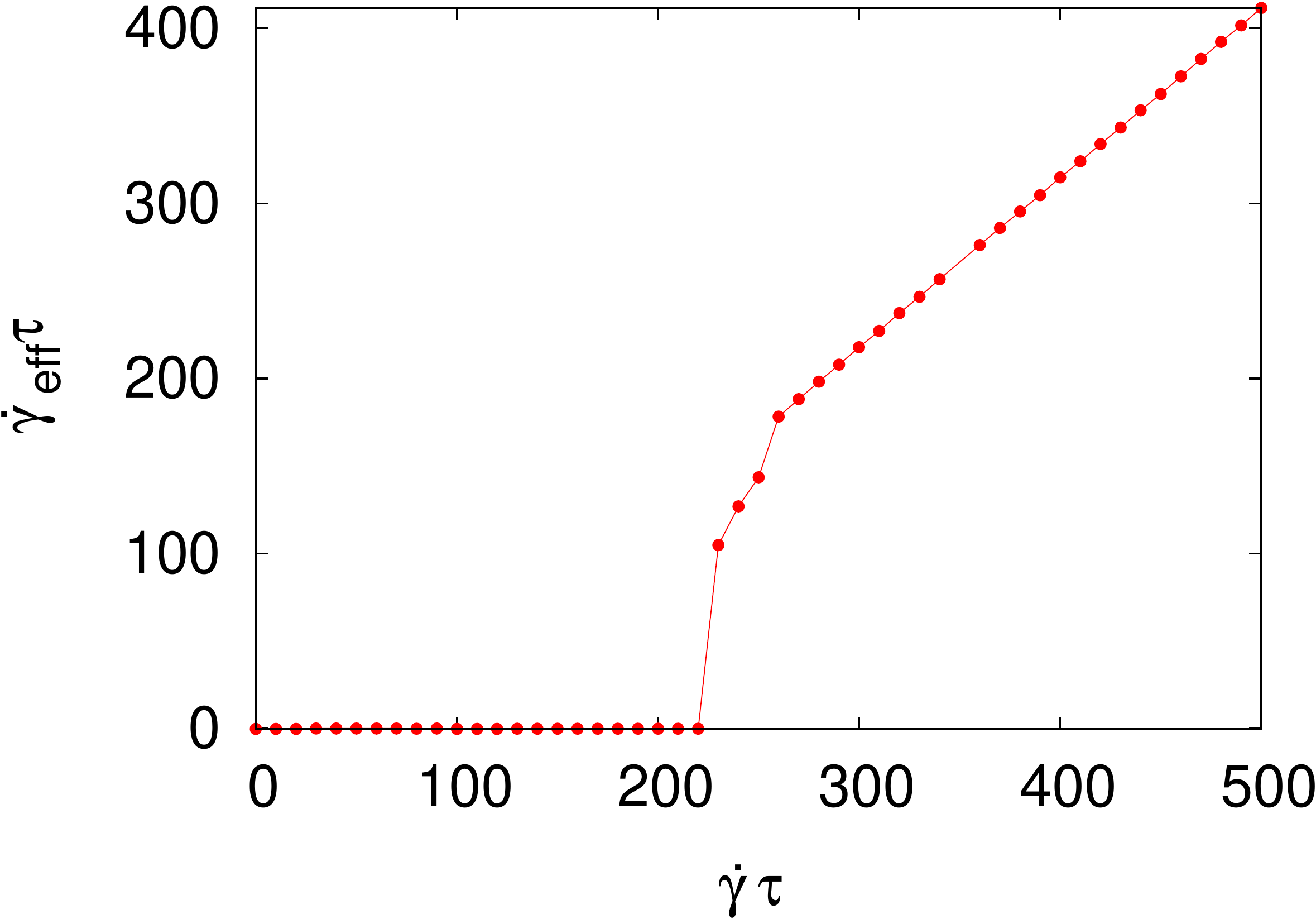}
\caption{(Color online) Effective shear rate (as determined from the velocity profile) as a function of the shear rate $\dot{\gamma}\tau$.}
\label{FIG:Veleff}
\end{figure}
Interestingly, the slope of these fitted linear profiles, $\dot{\gamma}_{eff}$, turns out to be different from the externally imposed value $\dot{\gamma}$. In order to understand the deviation between the applied and the effective shear rate we have plotted the effective shear rate $\dot{\gamma}_{eff}\tau$ as a function of the applied shear rate $\dot{\gamma}\tau$. The results are plotted in Fig.~\ref{FIG:Veleff}. Indeed $\dot{\gamma}_{eff}\tau \approx 0$ in the square-ordered regime ($\dot{\gamma}\tau\!\!\leq\!\!220$), reflecting the zero net motion of the particles. In the shear-melted and hexagonal state, we find $\dot{\gamma}_{eff}\tau>0$, but with values which are significantly lower than $\dot{\gamma}$. We conclude that the effective repulsive interaction between the layers (which is also responsible for the larger layer distance, see Fig.~\ref{FIG:Density} and thus, the increased friction, reduces the velocity of the layers.\par
\begin{figure}
\includegraphics[width=\linewidth]{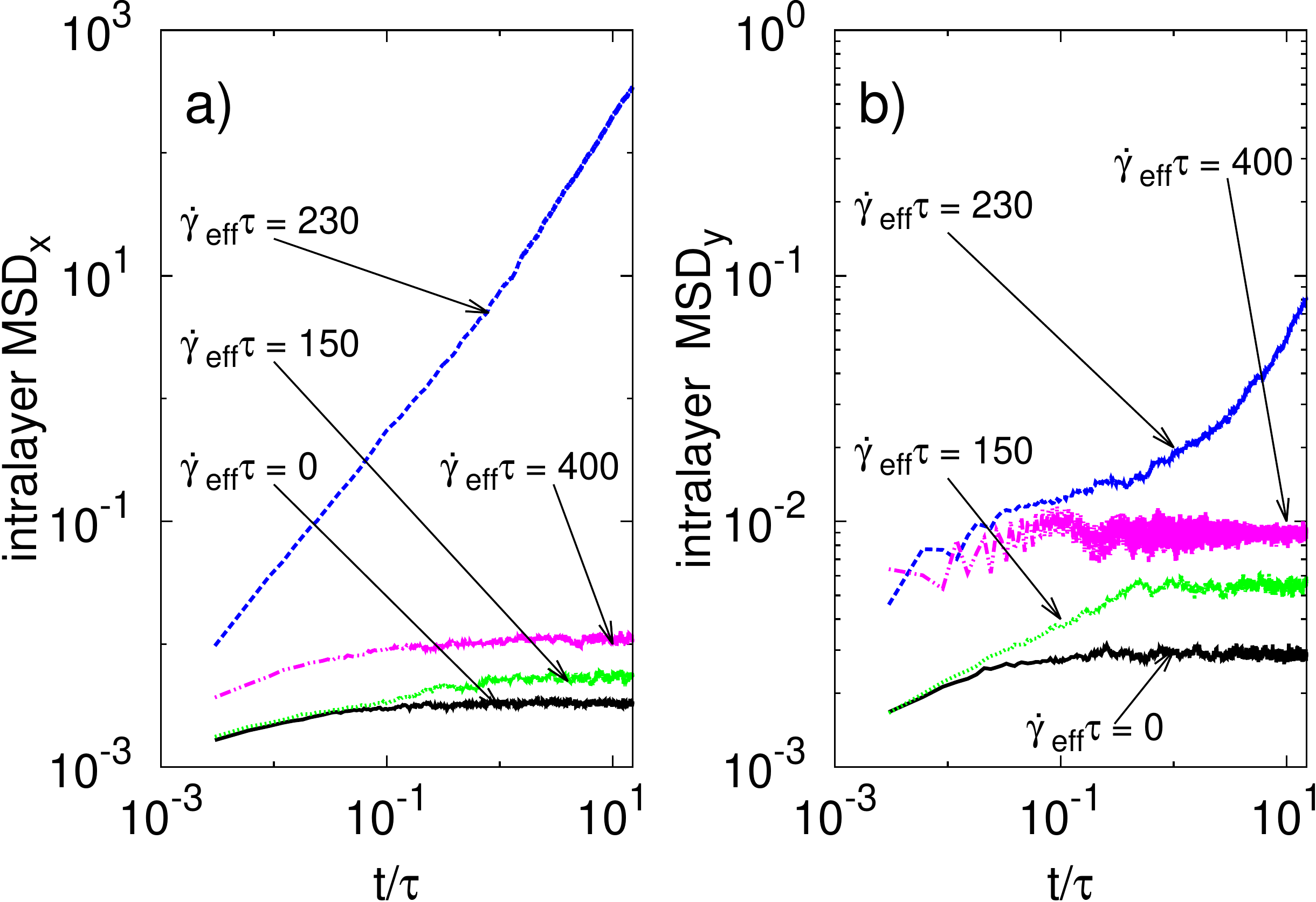}
\caption{(Color online) The position of the center of mass in flow direction (PCM$_{x}$) and the mean squared displacements within the layer in flow (MSD$_{x}$) and vorticity (MSD$_{y}$) directions at different shear rates.}
\label{FIG:MSD}
\end{figure}

Next, we consider the relative mean squared displacement (MSD) of particles within the layers. This function corresponds to the usual MSD corrected by the motion of the center of mass. Specifically,
\begin{align}
\Delta\mathbf{r}^{2}(t)=\left\langle\frac{1}{N_{l}}\sum_{i=1}^{N_{l}} \left(\mathbf{r}_{i}(t+t')-\mathbf{r}_{i}(t')-\Delta\mathbf{r}_{cm}(t)\right)^{2}\right\rangle.
\label{EQ:MSD}
\end{align}
By considering the relative MSD we can directly compare the mobility in the hexagonal and shear-melted state, where the entire layers move, with the mobility in the square-ordered state, where there is no such motion. Results for the components of the relative MSD in flow and vorticity direction are shown in Fig.~\ref{FIG:MSD}. At zero shear rate, both MSD components approach a plateau at long  times, reflecting the (persistent) trapping of the particles at the sites of the square lattice. \par
With the onset of the shear flow (small shear rates) this behavior at first persists. In fact, the y-component does indicate a small degree of mobility, but the entire behavior remains subdiffusive (i.e. $MSD\propto t^{\alpha}$ with $\alpha<1$) for the whole time range investigated. An appreciable in-plane mobility is only observed in the shear-melted state. Here, the particles escape their lattice positions and start to diffuse along both, the flow and the vorticity direction. The resulting non-trivial time dependence of the MSD will be analyzed in detail in Sec.~\ref{SEC:DT}. Upon further increase of $\dot{\gamma}$ we find from Fig.~\ref{FIG:MSD}a) that the MSD in flow direction develops a new plateau, thus indicating the reentrance of crystalline (this time, hexagonal) ordering within the still-moving layers. The emergence of a plateau at long times is also visible in the MSD in vorticity (i.e., $y$-) direction. 
\subsection{Diffusion in the shear-melted state \label{SEC:DT}}
In this section we attempt to describe the translational dynamics within the shear-melted state in an analytical fashion. The key ingredient is the fact that, although the lateral structure within this state is liquid-like, the particles are still essentially trapped in their layers. Here we approximate this confinement by a harmonic potential acting on each particle in $z$-direction, i.e. the shear gradient direction. We also assume that the confinement into the layers already represents the most important many-particle effect in our system. In other words, after introducing the (harmonic) confinement we consider the particles as independent. The corresponding equations of motion read
\begin{subequations}
\begin{align}
\dot{x}(t)&= \dot{\gamma}_{h} z(t) + \sqrt{2D_{h}}R_{x}(t)\\
\dot{y}(t)&=                     \sqrt{2D_{h}}R_{y}(t)\\
\dot{z}(t)&=- \frac{D_{h}\omega}{k_{B}T}         (z(t)-z_{0}) + \sqrt{2D_{h}}R_{z}(t)
\label{EQ:Trap}
\end{align}
\label{EQ:EqmotTrap1}
\end{subequations}
with $\langle R_{i}(t) \rangle=0$ and $\langle R_{i}(t)R_{j}(t') \rangle=\delta_{ij}\delta(t-t')$. The harmonic trap enters into \eq{EQ:Trap}, with $\omega$ playing the role of a spring constant. Similar equations have been recently derived and analyzed in \cite{Zimmermann10,Loewen12}. Solving \eq{EQ:EqmotTrap1} we obtain 
\begin{widetext}
\begin{subequations}
\begin{align}
x(t)&= x_{0}+                    \int_{0}^{t}\left(\dot{\gamma}_{h} z(t')+\sqrt{2D_{h}}R_{x}(t')\right)dt',\\
y(t)&= y_{0}+\sqrt{2D_{h}}           \int_{0}^{t}R_{y}(t')dt',\\
z(t)&= z_{0}+\sqrt{2D_{h}}\exp{\left(-\frac{D_{h}\omega}{k_{B}T}t\right)}\int_{0}^{t}\exp{\left(\frac{D_{h}\omega}{k_{B}T}t'\right)}R_{z}(t')dt'.
\end{align}
\label{EQ:EqmotTrap2}
\end{subequations}
\end{widetext}
From \eq{EQ:EqmotTrap2} we can calculate the MSDs in the flow direction,
\begin{widetext}
\begin{subequations}
\label{EQ:EqmotTrap3}
\begin{align}
\left \langle (x(t)-x_{0})^{2} \right \rangle &=\langle \int_{0}^{t}\left(\dot{\gamma}_{h} z(t')+\sqrt{2D_{h}}R_{x}(t)\right)dt'\int_{0}^{t}\left(\dot{\gamma}_{h} z(t'')+\sqrt{2D_{h}}R_{x}(t')\right)dt''\rangle \\
&= \int_{0}^{t} \int_{0}^{t} \dot{\gamma}^{2}_{h} \langle z(t') z(t'') \rangle dt' dt'' +2D_{h}t \\
&=\dot{\gamma}^{2}_{h}z_{0}^{2}t^{2}- \frac{\dot{\gamma}^{2}_{h}k_{B}^{3}T^{3}}{D^{2}_{h}\omega^{3}} \left(3+\exp{\left(-\frac{2 D_{h}\omega t}{k_{B}T}\right)}-4\exp{\left(-\frac{D_{h}\omega t}{k_{B}T}\right)}-\frac{2D_{h}\omega t}{k_{B}T} \right)+2D_{h}t\\
&=2D_{h}t+\dot{\gamma}^{2}_{h}z_{0}^{2}t^{2}+\frac{2 D_{h}\dot{\gamma}^{2}t^3}{3} + \dot{\gamma}^{2}_{h}D_{h} \left(\sum^{\infty}_{n=4}\frac{2^n t^n}{n!}\left(-\frac{D_{h}\omega }{k_{B}T}\right)^{n-3}-4\sum^{\infty}_{n=4}\frac{t^n}{n!}\left(-\frac{D_{h}\omega }{k_{B}T}\right)^{n-3}     \right) \label{EQ:EqmotTrap34}
\end{align}
\end{subequations}
\end{widetext}
and vorticity direction,
\begin{subequations}
\label{EQ:EqmotTrap4}
\begin{align}
\left\langle (y(t)-y_{0})^{2} \right\rangle &=2D_{h}\langle \int_{0}^{t}R_{y}(t)dt'\int_{0}^{t}R_{y}(t')dt''\rangle  \\
&= 2D_{h}t. \label{EQ:EqmotTrap42}
\end{align}
\end{subequations}
Inspecting \cref{EQ:EqmotTrap3,EQ:EqmotTrap4} we see that, within our idealized model, the MSD in vorticity direction displays normal diffusive behavior $\propto t$. Regarding the MSD in flow direction, the first and the second term on the right side of \cref{EQ:EqmotTrap34} represent contributions from the free diffusion and the net motion of the entire layer (with velocity $\dot{\gamma}z_{0}$), respectively. The third term $\propto t^{3}$ also appears when one considers a free particle under shear flow \cite{Krueger11}. It results from the diffusion of the particle along the velocity gradient. The last terms stem from the interplay of the harmonic confinement and the shear flow. Indeed, one finds that for $\omega\rightarrow0$ the MSD reduces to that for a free particle in shear flow. \par
\begin{figure}
\includegraphics[width=\linewidth]{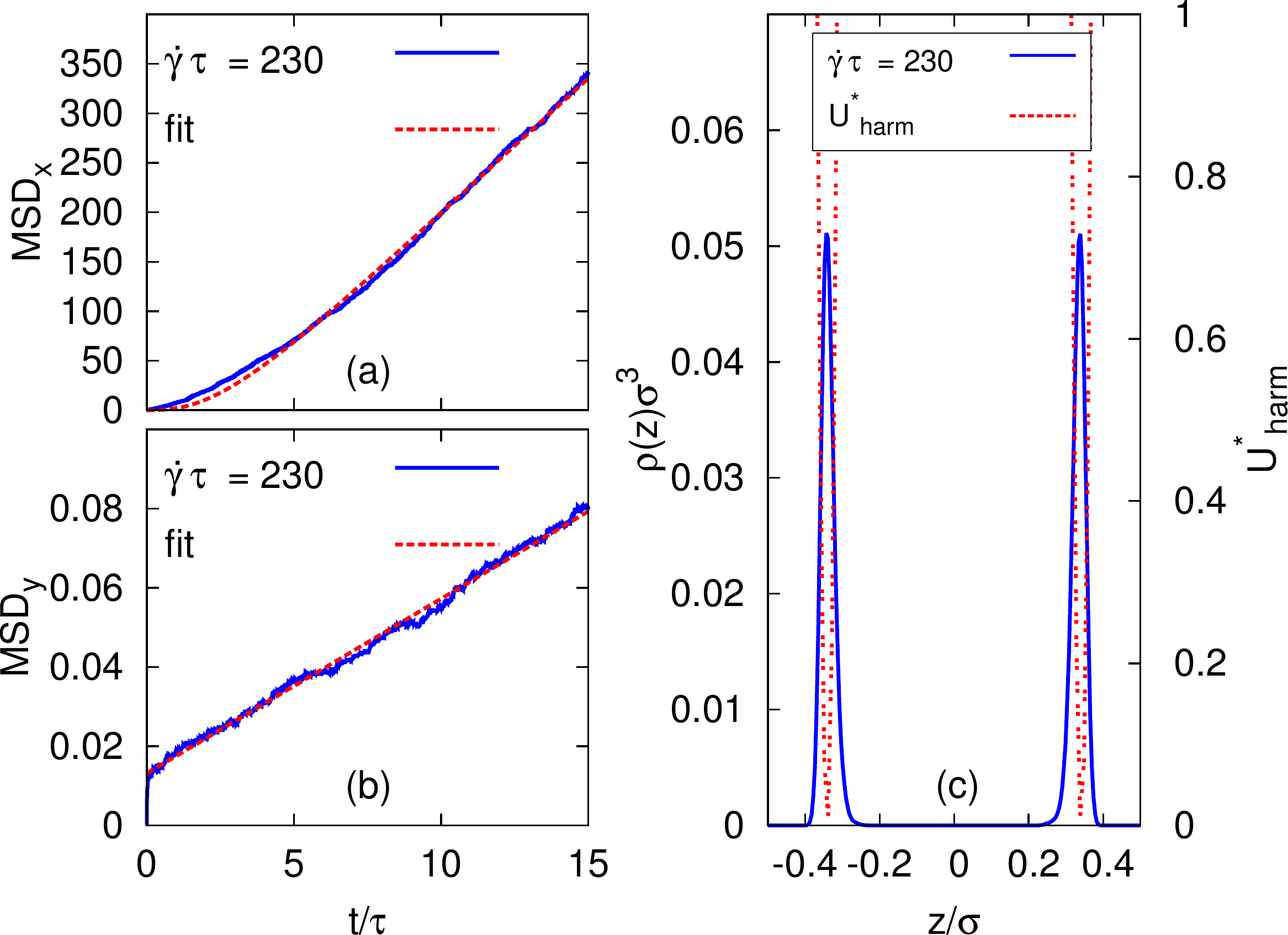}
\caption{(Color online) Fit of the mean squared displacements of one sheared particle in harmonic potential with the simulated system at $\dot{\gamma}\tau=230$ in flow direction (a) and vorticity direction (b). From the fit resulting harmonic potential was compared with the density profile in (c). The parameters used in the fit are $D_{h}^{x}=0.0004$, $D_{h}^{y}=0.002$ and $\omega=1419$.}
\label{FIG:Fit}
\end{figure}

To compare these analytical predictions with our simulation results, we set $k_{B}T$=1, $z_{0}$=$\pm$0.341 and $\dot{\gamma}_{h}$=104. The latter value corresponds to the effective shear rate at $\dot{\gamma}\tau=230$ [see~Fig.~\ref{FIG:Velocity}b)]. We then use the functions defined in \cref{EQ:EqmotTrap34,EQ:EqmotTrap42} to fit the simulation data, the fitting parameters being $D_{h}$ and $\omega$. The results of this procedure are shown in Fig.~\ref{FIG:Fit}a)-b), where we focus on the MSDs at long times. It turns out that the two MSDs obtained from the BD simulation can indeed be fitted nearly perfectly by the analytical expressions if we allow the diffusion constants $D_{h}$ to be different in $x$- and $y$-direction. Specifically, we find $D_{h}^{x}=0.0004$ and $D_{h}^{y}=0.002$. We understand these small values as a consequence of our strategy to describe the diffusion of a particle in a very dense, highly correlated system in an effective single-particle manner. The value obtained for the spring 
constant is $\omega=1419$. The resulting harmonic potential is plotted in Fig.~\ref{FIG:Fit}c) together with the density profile, showing that the widths of the harmonic wall describes that of the density profile very well.\par
Taken together, we can conclude that lateral diffusion in the shear-melted state of our confined system is indeed accurately described by the dynamics of a harmonically trapped free particle under flow. 
\subsection{Zig-zag motion at high shear rates\label{SEC:ZM}}
In this section we explore the appearance and characteristics of collective particle motion in our shear-driven colloidal bilayers. The presence of such collective modes is suggested by recent real-space experiments on three-dimensional colloidal crystals of Polymethylmethacrylat (PMMA) spheres under planar shear \cite{Derks09}. At suitable shear rates and densities, this real system consists of hexagonally ordered sliding layers, similar to what we see in our bilayer system. Investigating then the motion of individual PMMA spheres, a collective zig-zag motion into the vorticity direction (i.e., the $y$-direction) was observed \cite{Derks09}. In other words, strings of particles aligned in $x$-(flow) direction oscillate in phase. The microscopic origin of these oscillations can be understood when we recall the mutual arrangement of two hexagonal layers [see~Fig.~\ref{FIG:Snapshot}d)]. If these layers are moved relative to one another, each particle of one layer experiences barriers induced by the particles in 
the neighboring string in the adjacent layers. To circumvent these barriers the particles have to bypass and then come back, yielding effectively a zig-zag motion.\par
\begin{figure}
\includegraphics[width=\linewidth]{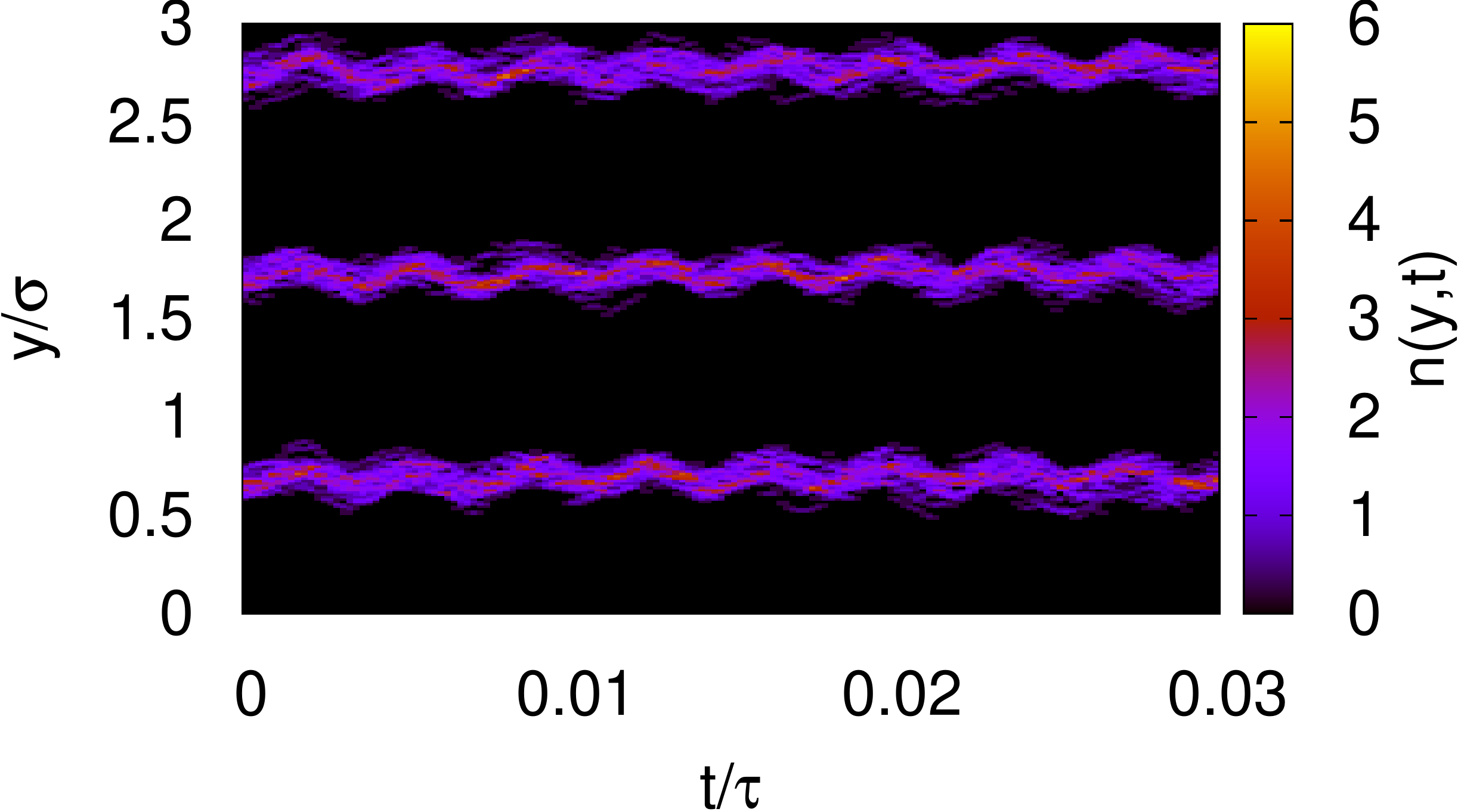}
\caption{(Color online) Time evolution of the number density $n(y,t)$ at $\dot{\gamma}~=~500$.}
\label{FIG:Zigsnap}
\end{figure}

We have found that the same kind of collective motion also occurs in our bilayer system. To visualize the zig-zag mode we plot in Fig.~\ref{FIG:Zigsnap} the number density of one layer in $y$-direction defined by 
\begin{align}
n(y,t)=\left \langle \sum_{i=1}^{N_{l}}\delta(y_{i}-y_{i}(t))\right \rangle
\end{align}
as function of time (horizontal axis) and space (vertical axis). One clearly identifies the regular character of the oscillations; also, neighboring strings apparently oscillate in phase.\par
\begin{figure}
\includegraphics[width=\linewidth]{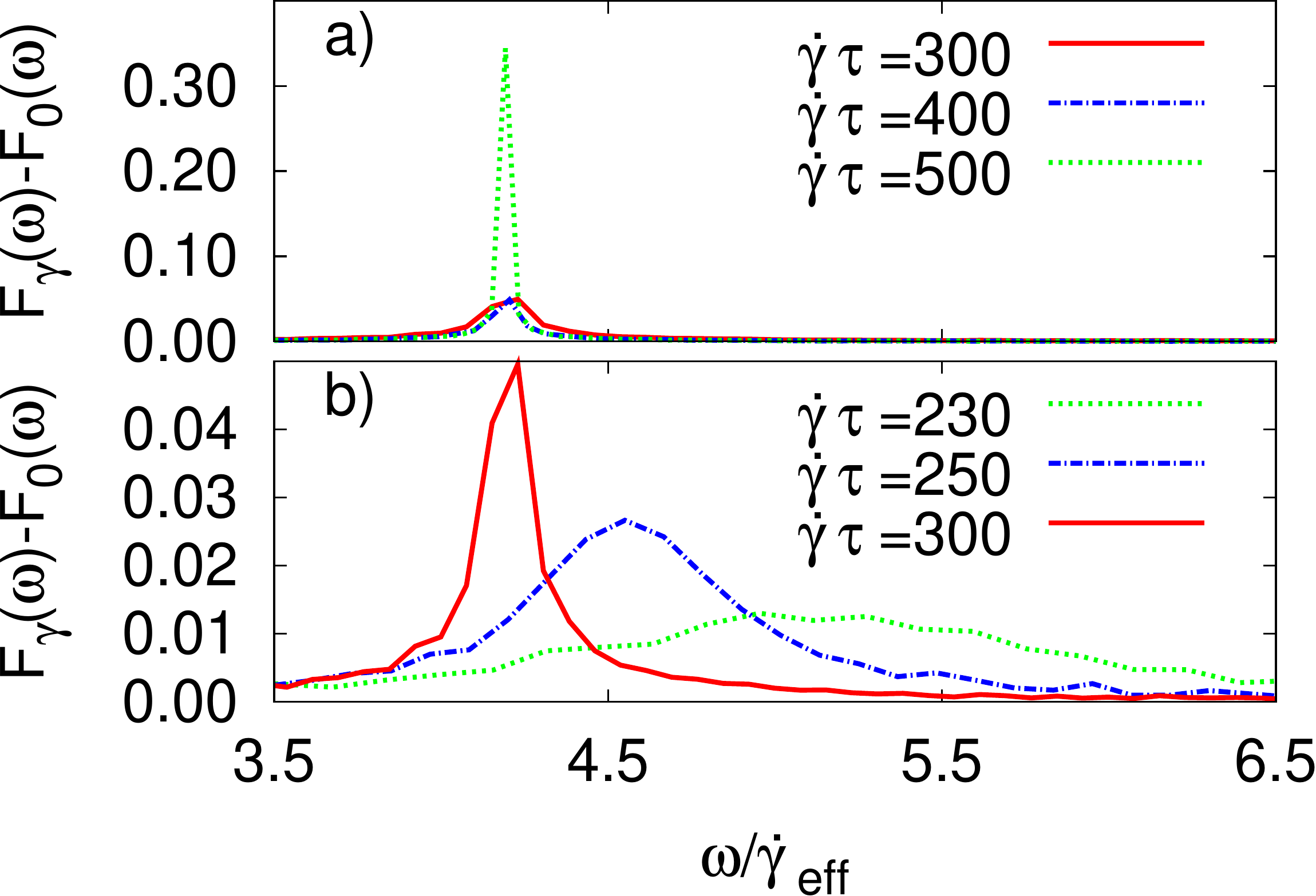}
\caption{(Color online) Rescaled oscillation frequencies of the zig-zag motion at different shear rates.}
\label{FIG:Zigzag}
\end{figure}
Having in mind the microscopic origin of the oscillations, one expects their frequency to increase with the shear rate. More precisely, the relevant shear rate in this context should be not $\dot{\gamma}$ (i.e., the externally imposed rate) but rather $\dot{\gamma}_{eff}$, which corresponds to the effective local shear rate already considered in Sec.~\ref{SEC:TM} [see~Fig.~\ref{FIG:Velocity}b)]. Indeed, in the experiments \cite{Derks09} it was found that the oscillation frequency $\omega_{0}$ is proportional to the effective shear rate. To test this expectation in our system, we now consider the Fourier transform of the center-of-mass position of one layer as function of time, 
\begin{align}
F_{\dot{\gamma}}(\omega)=\left\langle \int_{0}^{t_{max}} \Delta y_{cm}(t)\exp(-\omega t)dt \right\rangle,
\label{EQ:Fourier}
\end{align}
where $\Delta y_{cm}(t)$ describes the shift of the position of the center of mass as a function of time. Results for $F_{\dot{\gamma}}(\omega)$ at different shear rates are plotted in Fig.~\ref{FIG:Zigzag}. Note that we have scaled the frequency (horizontal axis) by the effective shear rate, $\dot{\gamma}_{eff}$. From Fig.~\ref{FIG:Zigzag}a) we see that, within the hexagonally ordered state ($\dot{\gamma}\tau \geq 300$), the functions $F_{\dot{\gamma}}(\omega)$ are essentially zero apart from one, relatively sharp peak, with its height and sharpness increasing with increasing $\dot{\gamma}$. This signals the presence of one dominant oscillation frequency, consistent with the real-time plot in Fig.~\ref{FIG:Zigsnap}. Moreover, the peaks are located at the {\it same} scaled frequency $\omega_{0}/\dot{\gamma}_{eff}\approx4.19$. This scaling is in agreement to the corresponding experimental observation \cite{Derks09}. \par
For the present system, we have found that the oscillation frequency $\omega_{0}$ obeys the relation
\begin{align}
\frac{\omega_{0}}{2\pi}=\frac{\dot{\gamma}_{eff} \Delta z}{x_{0}},
\label{EQ:Zigzag}
\end{align}
where $\Delta z=0.69\sigma$ is the distance between the layers and $x_{0}$ corresponds to the in-plane distance between nearest neighbors in flow direction. From the in-plane correlations plotted in Fig.~\ref{FIG:Intra} we obtain $x_{0}\approx1.03\sigma$. The constancy of the ratio $\omega_{0}/\dot{\gamma}_{eff}$ for a range of shear rates then shows that $\Delta z$ and $x_{0}$ remain approximately constant within the hexagonal regime. Figure~\ref{FIG:Zigzag}b) contains additional data for $F_{\dot{\gamma}}(\omega)$ at lower shear rates. Recall that the values $\dot{\gamma}\tau=230$ and 250 pertain to the shear -melted regime. As seen in Fig.~\ref{FIG:Zigzag}b), the corresponding functions $F_{\dot{\gamma}}(\omega)$ still have one maximum, but its height is much lower and the entire distribution is much broader than in the hexagonal regime. In other words, there are still oscillations in the shear-melted state, but these have a broader spectrum of frequencies and are less synchronized. Moreover, the location 
of the maximum is different from the  value $\omega_{0}/\dot{\gamma}_{eff}\approx4.19$ in the hexagonal state and now depends on $\dot{\gamma}$. This can be understood when we recall that the particles in the shear-melted regime are not bound to lattice sites, thus, there is no fixed nearest neighbor distance. For completeness, we also note that the maximum of $F_{\dot{\gamma}}(\omega)$ still present in the shear-melted state vanishes entirely upon further decrease of $\dot{\gamma}$ into the square-ordered state. Hence, there is no motion of the layers, and thus no need for the particles to perform oscillations.\par
In view of the collective zig-zag motion of the particles, seen in the number density $n(y,t)$ (or, equivalently, the center of mass) it is an interesting question whether spatial correlations between the particles are influenced as well. To this end we now consider the distinct part of the intralayer van Hove correlation function defined by 
\begin{align}
G(y,t)=\left\langle \frac{1}{N_{l}} \sum^{N_{l}}_{i=1}\sum^{N_{l}}_{j \neq i} \delta[y-y_{j}(t+t')+y_{i}(t')] \right\rangle.
\label{EQ:VHOVE}
\end{align}
\begin{figure}
\includegraphics[width=\linewidth]{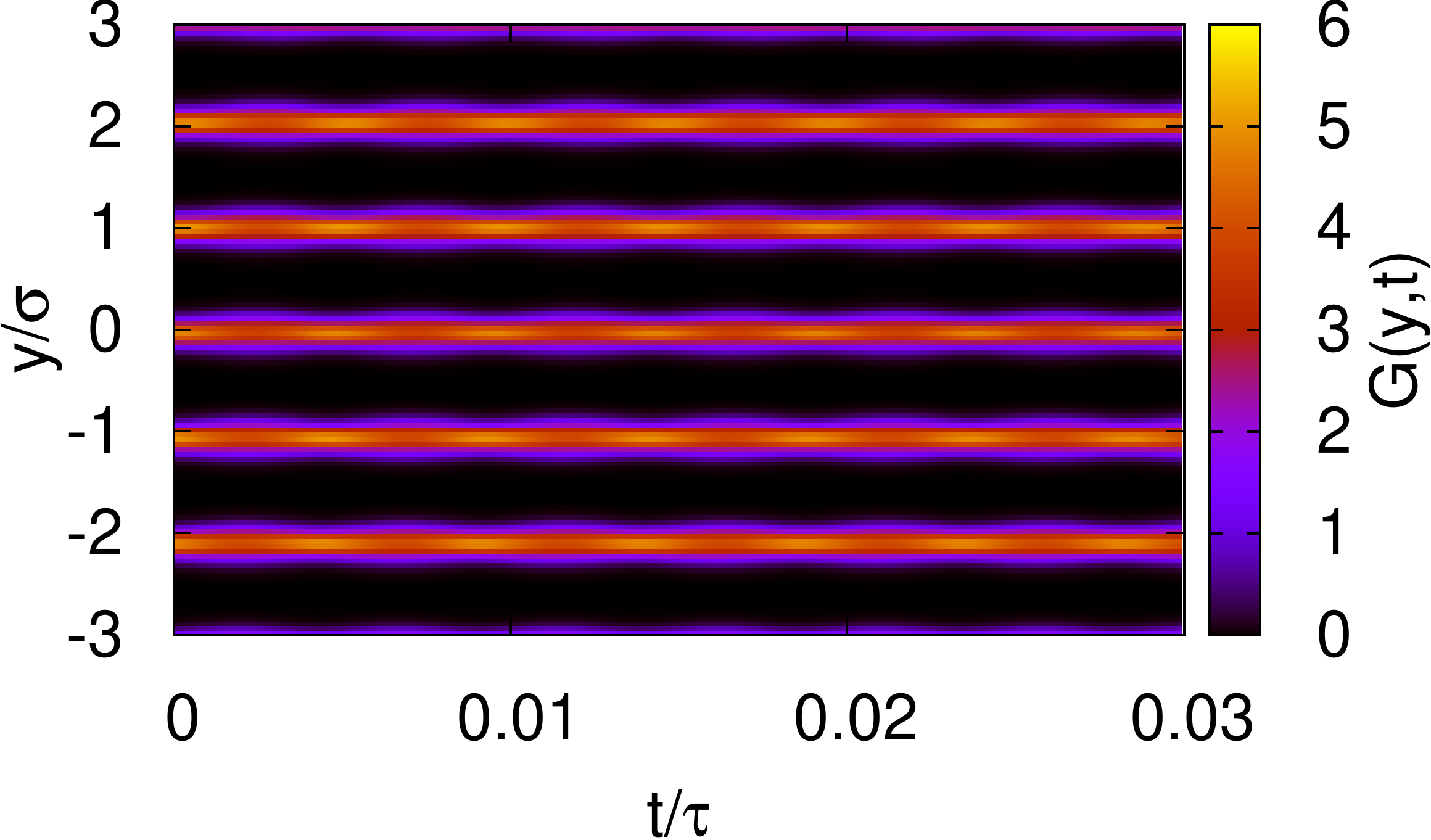}
\caption{(Color online) The distinct part of the van Hove function in vorticity direction at $\dot{\gamma}~=~400$.}
\label{FIG:VanHove}
\end{figure}
\begin{figure}
\includegraphics[width=\linewidth]{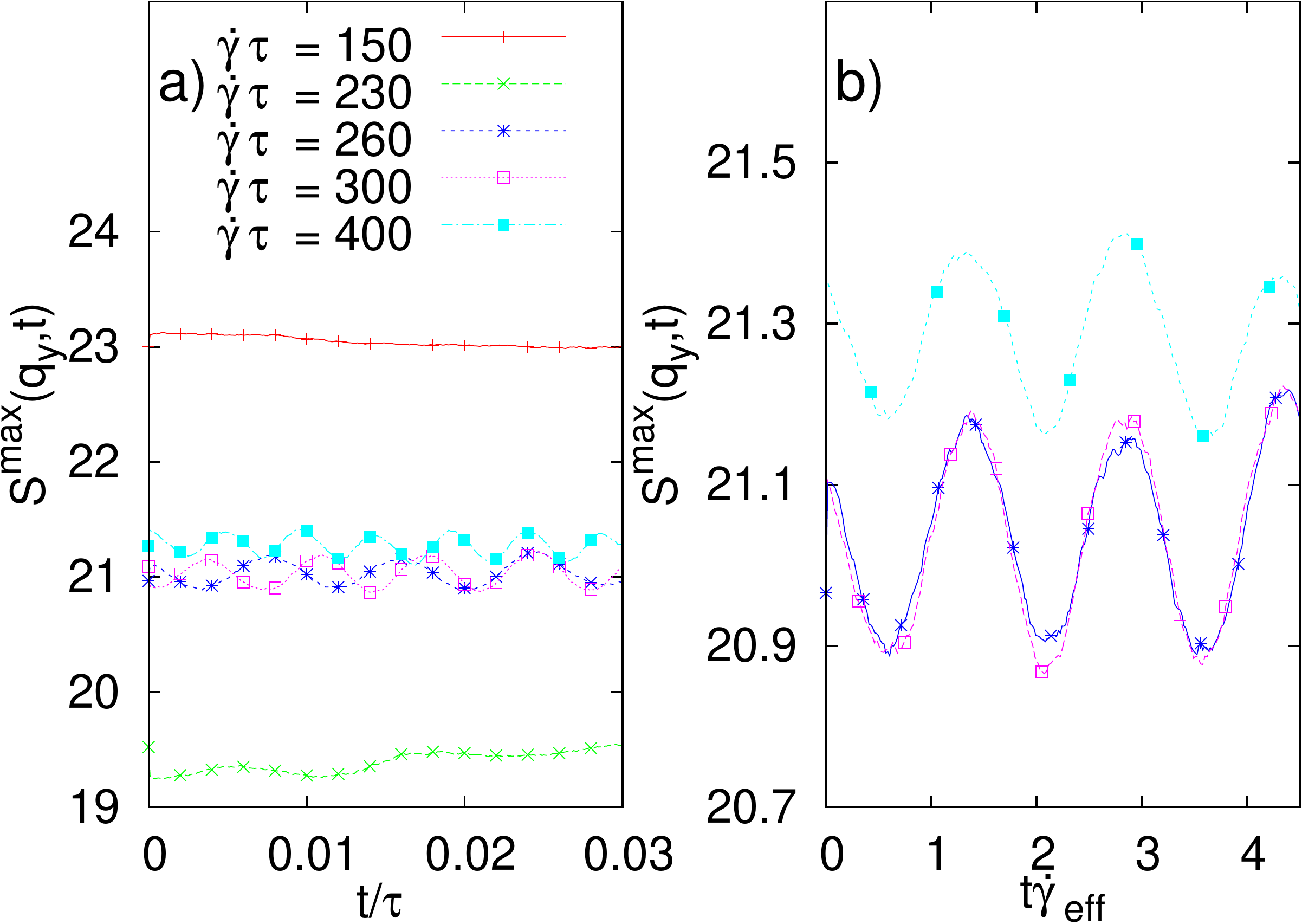}
\caption{(Color online) Rescaled oscillation of the structure factor in the vorticity direction.}
\label{FIG:Struct}
\end{figure}
At $t=0$ this function reduces to the conventional pair correlation function in $y$-direction. A representative result obtained at  $\dot{\gamma}\tau=400$ is shown in Fig.~\ref{FIG:VanHove}. The plot clearly reflects strong correlations induced by the hexagonal ordering. Note that, by definition, the oscillations of $n(y,t)$ are not directly visible in $G(y,t)$. However, closer inspection of the time dependence of the correlations at fixed $y$-distance reveals that there are oscillations in the {\it degree} of the correlations. The period of these oscillations appears to be constant over time. For a better analysis we have calculated the spatial Fourier transform of $G(y,t)$, yielding the dynamic structure factor in vorticity direction, $S(q_{y},t)$. As expected in a strongly correlated system, $S(q_{y},t)$ (not plotted here) displays pronounced maxima at multiple of $q=2\pi/y_{0}$ where $y_{0}\approx1\sigma$ is the average nearest neighbor distance in $y$-direction. The height of the largest peak, $S^{max}(
q_{y},t)$ can then be interpreted as a measure of the degree of correlations. In Fig.~\ref{FIG:Struct}a), we plot $S^{max}(q_{y},t)$ at different shear rates. We can easily see that in the crystalline square-ordered state the correlation level of the system is very high. Increasing the shear rate and entering in the melted state the correlation strength in the system decreases until the layers recrystallize in the hexagonal state which leads to the increase of $S^{max}(q_{y},t)$. Furthermore we can see, that the correlation strength in the hexagonal state performs oscillations. In order to understand this oscillatory behavior we plotted $S^{max}(q_{y},t)$ rescaling the timescale with  $\dot{\gamma}_{eff}$ for several shear rates pertaining to the hexagonal regime. The results are visualized in Fig.~\ref{FIG:Struct}b). One observes indeed regular oscillations, confirming that not only the particles themselves, but also their correlations oscillate in time. Moreover, the frequency of the correlational 
oscillations coincides with those obtained for the position of the center of mass [see~Fig.~\ref{FIG:Zigzag}]. We understand these correlations as follows: During one period of the zig-zag motion, a particle moves from its original position on the $y$-axis towards a position dictated by the ``bypassing'' of the barrier induced by particle in the adjacent layer, and then back to its original $y$-position. In the instant corresponding to the bypass, the available space is severely restricted. The same happens to the neighboring particle in the same layer in $y$-direction. As a consequence, correlations during the ``bypass'' are enhanced relative to those in the original arrangement.

%%%%%%%%%%%%%%%%%%%%%%%%%%%%%%%%%%%%%%%%%%%%%%%%%%%%%%%%%%%%%%%%%
%%%%%%%%%%%%%%%%%%%%%%%         CONCLUSION       %%%%%%%%%%%%%%%%%%%%%
%%%%%%%%%%%%%%%%%%%%%%%%%%%%%%%%%%%%%%%%%%%%%%%%%%%%%%%%%%%%%%%%%
\section{Conclusions \label{SEC:CONC}}
In this paper we have presented a BD simulation study of a bilayer composed of charged colloidal particles subject to planar shear flow. Starting from a high-density equilibrium state with lateral crystalline (square-like) order and switching on the shear flow, we find that the structure first remains unchanged, with both layers being "pinned" in their original position (i.e., zero net motion of the layers). Only above a certain shear rate the layers start to move. In this regime, the crystalline order first melts, followed by a recrystallization into a state with {\it hexagonal} in-plane ordering. We have then analyzed, in detail, the translational single-particle and collective motion within the shear-melted and hexagonal state. \par
 One key result of our study is that the time-dependence of the in-plane mean-squared displacements in the shear-melted (yet strongly correlated) state can be accurately described by a simple, analytically accessible model. This model involves a shear-driven {\it single} particle in a harmonic trap acting in shear gradient direction, the latter representing the effective restriction of the particle's motion within layers parallel to the confining walls. In the hexagonal state, on the contrary, the in-plane MSD's (corrected by the net layer motion) are characterized by plateaus reflecting the caging of the particles around the sites of the hexagonal lattice. \par
 Another main result of our study is that the sheared bilayer system in its recrystallized (hexagonal) state displays a "zig-zag" collective mode, characterized by oscillatory synchronized motion of strings of particles along the vorticity direction. The frequency of these oscillations scales with the effective shear rate. The same behavior has been recently observed in experiments of shear-driven colloidal films, yet at macroscopic film thicknesses \cite{Derks09}. Moreover, we have shown that the oscillations of the particle positions are accompanied by oscillations of the spatio-temporal correlation functions. To quantify this behavior we have investigated the height of the maximum of the time-dependent structure factor, which shows regular oscillations in the hexagonal (but not in the shear-melted) state. \par 
 At this point it seems worth to recall that our model calculations (including the chosen parameter set) pertain to a realistic system of charged silica particles \cite{Grandner08, Grandner09}. Thus, our predictions can in principle be tested by experiments. One main feature characterizing the real systems, but missing in our model, is the presence of a solvent, which would induce hydrodynamic interactions between the colloidal particles. However, previous simulation studies of sheared colloids have suggested that hydrodynamic interactions  mainly affect the timescales \cite{Besseling12} and tend to smoothen correlational effects but not alters the qualitative behavior. Thus, we are confident that our observations are detectable in real systems. \par
 Starting from the present study, there are several interesting routes to follow. An obvious extension is towards oscillatory shear flow, a situation which has also been investigated in experiments of 3D colloidal structures \cite{Besseling12} and in simulations \cite{Brader10, Besseling12}. Of particular importance would then be the (presumably highly nonlinear) rheological response of the system. Further, given the shear-induced changes of translational order observed  in the present study it would be interesting to see to which extent these ordering types survive when using {\it structured}, crystalline walls (see \cite{Wilms12} for  a corresponding recent BD study of a much thicker colloidal film). Finally, it seems worth to establish in more detail the links between the present systems, where the two layers  are sheared relative to one another, and models for nano friction, where one typically considers the motion of a crystalline monolayer of particles over a rough substrate. That there is, indeed, 
such as link is already suggested from the behavior of the effective shear rate as function of the true $\dot\gamma$ [see~Fig.~\ref{FIG:Veleff}]. This behavior is very similar to that observed for the effective velocity in driven monolayers \cite{Tosatti12,Hasnain13}. Moreover, for nano friction models (such as the Frenkel-Kontorova-model) it is very well established that they display complex dynamical behavior, an example being the appearance of kinks and antikinks \cite{Tosatti13}. Work in these directions is in progress.

%%%%%%%%%%%%%%%%%%%%%%%%%%%%%%%%%%%%%%%%%%%%%%%%%%%%%%%%%%%%%%%%%
%%%%%%%%%%%%%%%%%%%%%%%         CONCS        %%%%%%%%%%%%%%%%%%%%%
%%%%%%%%%%%%%%%%%%%%%%%%%%%%%%%%%%%%%%%%%%%%%%%%%%%%%%%%%%%%%%%%%
%%%%%%%%%%%%%%%%%%%%%%%%%%%%%
\begin{acknowledgments}
This work was supported by the Deutsche Forschungsgemeinschaft through SFB 910 (project B2).
\end{acknowledgments}
%%%%%%%%%%%%%%%%%%%%%%%%%%%%%%%%%%%%%%%%%%%%%%%%%%%%%%%%%%%%%%%%%
%%%%%%%%%%%%%%%%%%%%%%%%%%%%%%%%%%%%%%%%%%%%%%%%%%%%%%%%%%%%%%%%%%%
%%%%%%%%%%%%%%%%%%%%%%%%%%%%%%%%%%%%%%%%%%%%%%%%%%%%%%%%%%%%%%%%%%%%%

\end{document}